\documentclass{article}
\usepackage[]{geometry}
\newgeometry{
  textheight=9in,
  textwidth=5.5in,
  top=1in,
  headheight=12pt,
  headsep=25pt,
  footskip=30pt
}
\usepackage[normalem]{ulem}
\usepackage{xcolor}
\usepackage{amsmath, amssymb}
\usepackage{graphicx}
\usepackage[final]{pdfpages}
\usepackage{subcaption}
\usepackage{algorithm}
\usepackage{algpseudocode}
\usepackage{url}
\usepackage{bm}
\usepackage{authblk} 

\title{$\text{AB}\mathbb{C}\text{MB}$: Deep Delensing Assisted Likelihood-Free Inference from CMB Polarization Maps}
\date{}
\begin{document}
\author[1]{Kai Yi\thanks{Corresponding author: kai.yi@unsw.edu.au}}
\author[1,2]{Yanan Fan}
\author[3]{Jan Hamann}
\author[4]{Pietro Li\`{o}} 
\author[1,5]{Yu Guang Wang}

\affil[1]{School of Mathematics and Statistics, UNSW  Sydney, NSW 2052, Australia}
\affil[2]{Data61, CSIRO,  Sydney NSW 2015, Australia}
\affil[3]{School of Physics, UNSW Sydney, NSW 2052, Australia}
\affil[4]{Computer Laboratory, University of Cambridge\\ Cambridge CB3 0FD, UK} 
\affil[5]{Institute of Natural Sciences, Shanghai Jiao Tong University\\ Shanghai 200240, China}

\maketitle

\begin{abstract}

The existence of a cosmic background of primordial gravitational waves (PGWB) is a robust prediction of inflationary cosmology, but it has so far evaded discovery.  The most promising avenue of its detection is via measurements of Cosmic Microwave Background (CMB) $B$-polarization.  However, this is not straightforward due to (a) the fact that CMB maps are distorted by gravitational lensing and (b) the high-dimensional nature of CMB data, which renders likelihood-based analysis methods computationally extremely expensive.  In this paper, we introduce an efficient likelihood-free, end-to-end inference method to directly infer the posterior distribution of the tensor-to-scalar ratio $r$ from lensed maps of the Stokes $Q$ and $U$ polarization parameters. Our method employs a generative model to delense the maps and utilizes the Approximate Bayesian Computation (ABC) algorithm to sample $r$.  We demonstrate that our method yields unbiased estimates of $r$ with well-calibrated uncertainty quantification.
\end{abstract}

\section{Introduction}

The Cosmic Microwave Background (CMB) – emitted about 13.8 billion years ago – is the oldest light in the Universe, and measurements of its anisotropies form the basis of much of our knowledge about cosmology today~\cite{Planck:2018nkj}.  While the CMB's temperature fluctuations have been measured with great accuracy by the \textit{Planck} mission, there is still a lot of untapped information in the polarization of the CMB, which will be the focus of upcoming experiments like CMB-S4~\cite{CMB-S4:2022ght} and LiteBIRD~\cite{LiteBIRD:2020khw}.  The CMB's polarization signal has two degrees of freedom and can be expressed in terms of Stokes $Q$ and $U$ parameters, or, alternatively, the curl-free $E$ mode and the divergence-free $B$ mode.  The $B$ mode, in particular, turns out to be a particularly sensitive observable for detecting signatures left behind by background of primordial gravitational waves whose existence is generically predicted in models of cosmic inflation~\cite{Starobinsky:1979ty}.  A discovery of an inflationary PGWB would therefore open a window on the physics that governs the very first moments of our Universe, at energies exceeding those testable in terrestrial laboratories by many orders of magnitude.

At lowest order, the PWGB can be parameterized by the tensor-to-scalar ratio $r$, i.e., the amplitude of the PGWB's (tensor) power spectrum relative to that of density (scalar) perturbations – and $r$ is directly proportional to the energy scale of inflation.  Current CMB experiments have placed an upper limit on $r$ (\textit{Planck} alone gives $r  < 0.10$ at  95\% confidence level \cite{Planck:2018vyg}, and in combination with BICEP-Keck this reduces to $r < 0.036$~\cite{BICEPKeck:2022mhb}), and already ruled out certain classes of inflationary models~\cite{Planck:2018jri}.  Within the next decade, a sensitivity of $\sigma(r) \sim 10^{-3}$ is targeted.

Reaching this goal is complicated by the fact that an additional $B$ mode signal is induced by weak gravitational lensing of the CMB, which effectively acts as a contaminant to the primordial $B$ mode polarization due to the PGWB. The ability to remove at least some of this lensing contribution – \textit{delensing} – will thus be important if one wants to maximize the discovery potential of upcoming experiments. 
The standard method for delensing is the use of a quadratic estimator (QE)~\cite{Hu:2001kj}, however this becomes suboptimal for CMB noise levels below a few $\mu\text{K-arcmin}$~\cite{Seljak:2003pn}, achievable by next-generation experiments.
More efficient approaches to the delensing problem have been considered in the literature, e.g., via neural networks~\cite{caldeira2019deepcmb,heinrich2024cmb,yi2022approximate} – which have the advantage that once the neural network is trained, the actual delensing procedure does not incur huge computational cost.  However, in these previous works $r$ is assumed to be known and these delensing techniques may thus not be suitable for inferring $r$ in the usual way, i.e., by constructing a likelihood and then performing posterior inference using Markov chain Monte Carlo (MCMC), see e.g., \cite{BICEP:2021xfz, Eriksen_2008, millea2019bayesian}.  

A very promising alternative was considered by Millea et.al.~\cite{Millea:2020cpw}, who introduced the idea to sample the full joint posterior probability of the cosmological parameters and the unlensed maps.  This naturally allows the inference of $r$, but can be computationally demanding due to the high number of dimensions involved.

\begin{figure}[!htp]
    \centering
    \includegraphics[width=\textwidth, height=4cm]{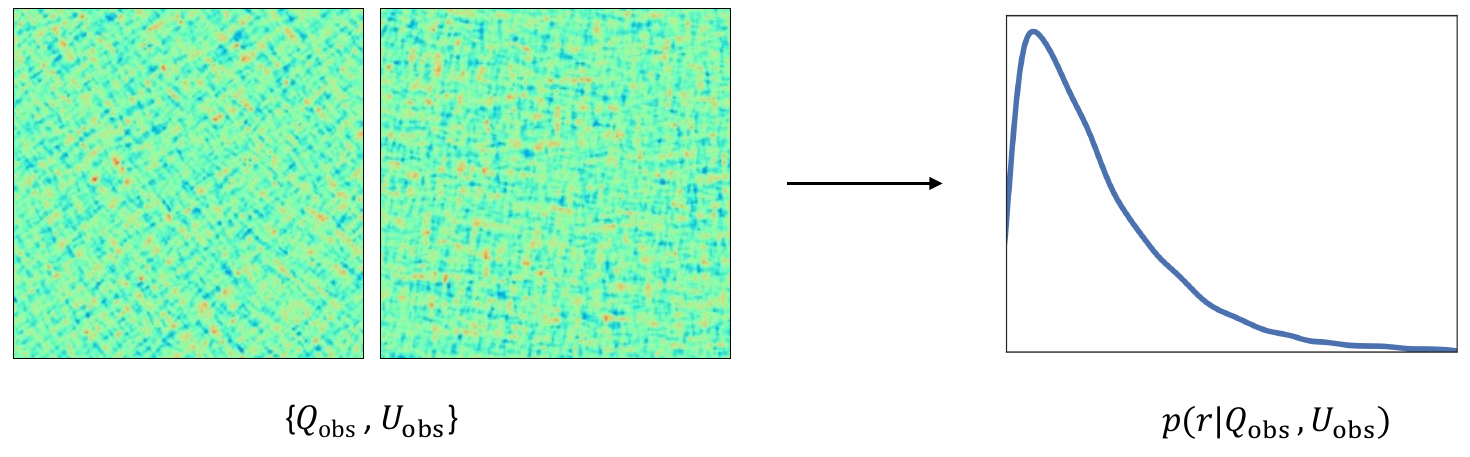}
    \caption{Illustration of the process from input to output.  Maps of the $Q$ and $U$ polarization parameters observed over a $12 \times 12$ square degree patch (left) serve as input, and posterior distribution of the tensor-to-scalar ratio $r$ (right) as output.
}
    \label{fig:motivation}
\end{figure}

In this paper, we introduce a Bayesian inference framework that utilizes a likelihood-free approach (also known as Approximate Bayesian Computation (ABC)~\cite{sisson2018handbook}), based on data simulations~\cite{grazian2019}, to infer the posterior distribution of the tensor-to-scalar ratio $r$ from lensed polarization maps $Q_\mathrm{obs}$ and $U_\mathrm{obs}$ (as shown in Figure \ref{fig:motivation}).  Informed by the underlying physics, our method incorporates a generative neural network model as an intermediate step (not shown in Figure \ref{fig:motivation}) for estimating a delensed $B$ polarization map whose power spectrum is then used as a summary statistic in the subsequent ABC inference step.  Importantly, the neural network is trained with simulations covering a range of values of $r$ and we therefore do not need to impose a fixed value of $r$ for the delensing procedure.  We show that our approach produces unbiased estimates of $r$ with well calibrated uncertainty quantification.

The paper is structured as follows: in Sections~\ref{sec:abc} and~\ref{sec:cmb} we briefly recount the basics of ABC and the CMB observables, respectively.  We introduce our delensing algorithm in Section~\ref{sec:VAE} and discuss the detailed implementation of ABC inference in Section~\ref{sec:abcinference}.  Our results are presented in Section~\ref{sec:results} and we conclude in Section~\ref{sec:conclusions}.

\section{Bayesian Inference using Simulator Model \label{sec:abc}}

Approximate Bayesian Computation (ABC) is a class of algorithms which have been developed to perform Bayesian inference in the absence of a likelihood function, see \cite{sissonyb2018} for an overview. A defining feature of these algorithms is the reliance on a known data generating mechanism, so that for any value of the model parameter, $\theta$, we can obtain pseudo-observations using the same mechanism that generated the observed data. For example, in CMB analysis, $\theta$ may represent the lensing potential parameter $\phi$ and/or the tensor-to-scalar ratio $r$, and data or images can be generated from a cosmological model.
This is sometimes referred to as a simulator model, i.e. models which are specified only through the generative mechanism, where the likelihood function is not required explicitly in analytic form. Examples of applications of ABC can typically be found in problems where the likelihood is either too expensive to compute or difficult to specify.

Bayes' theorem dictates that the posterior distribution for the tensor-to-scalar ratio $r$ is given by
\[
p(r|Q_\mathrm{obs}, U_\mathrm{obs}) \propto L(Q_\mathrm{obs}, U_\mathrm{obs}|r) p(r)
\]
where $L(Q_\mathrm{obs}, U_\mathrm{obs}|r)$ denotes the likelihood function for the observed data $Q_\mathrm{obs}, U_\mathrm{obs}$, and $p(r)$ is the prior for $r$. Then a rejection sampling based ABC algorithm for approximating $p(r| Q_\mathrm{obs}, U_\mathrm{obs})$ proceeds by sampling $r_n$,  $n=1,\ldots, N$, from the prior distribution $p(r)$,  and for each $r_n$, $Q_n, U_n$ maps are sampled via the simulator $p(Q, U|r_n)$. If the sampled $Q_n, U_n$ maps are within a small $\epsilon >0 $ of the observed $Q_\mathrm{obs}, U_\mathrm{obs}$ maps, the sample $r_n$ is retained as a posterior sample. We summarize the algorithm as follows.

\begin{minipage}{0.95\textwidth}
    \begin{algorithm}[H]
        \caption{Vanilla Approximate Bayesian Computation (ABC) Algorithm}  
        \begin{algorithmic}[1]
            \For{$n = 1 : N$}
                \State Sample $r_n \sim p(r)$.
                \State Simulate maps $Q_n, U_n \sim p(Q, U | r_n)$ using the simulator $p(Q, U|r_n)$, and 
                compute summary statistics $S(Q_n, U_n)$.
                \State Accept $r_n$ if $\|S(Q_n, U_n) - S(Q_\mathrm{obs}, U_\mathrm{obs})  \|_2  < \epsilon$.
            \EndFor

        \end{algorithmic}  
        \label{vABC algorithm}
    \end{algorithm}
\end{minipage}

It is common to measure the discrepancy $\|(Q_n, U_n) - (Q_\mathrm{obs}, U_\mathrm{obs})  \|_2 $  between the high-dimensional observed and simulated data via a low dimensional set of summary statistics $S(Q_n, U_n)$ and $S(Q_\mathrm{obs}, U_\mathrm{obs})$. Here, we will employ neural networks to reconstruct unlensed $B$ maps and use their power spectrum as summary statistics.
Further details on the summary statistics calculation will be discussed in Section \ref{sec:VAE}. 
In the following sections, we will describe our approach to each of the components required in ABC.

\section{CMB Polarization and the Data Generation Process \label{sec:cmb}}
In this section, we describe the data generation process and provide details on the datasets used in the paper.
%
%
We first describe the lensing effect, and introduce the generation of unlensed $B$ maps via a physical model as an intermediate step before obtaining the $Q$ and $U$ maps via a lensing algorithm.


\subsection{CMB Observables and the Lensing Effect}

Although the CMB's polarization signal is significantly fainter than its intensity,
the WMAP ~\cite{WMAP:2012fli} and Planck~\cite{Planck:2018nkj} missions have produced full-sky maps of the CMB polarization at a modest signal-to-noise ratio.  In addition to this, ground-based experiments such as BICEP2/Keck \cite{BICEP2:2015vut}, SPT~\cite{SPT:2019nip}, or ACTPol~\cite{ACTPol:2014pbf} have been able to provide higher signal-to-noise measurements in localised, low-foreground regions of the sky.

The polarization of the CMB has two degrees of freedom that are typically represented in one of two bases: (i) the Stokes parameters $(Q,U)$ which depend on the spherical coordinate system chosen (but are directly related to what a polarization detector measures), and (ii), the $(E,B)$ basis, with a gradient-like $E$ component and a curl-like $B$ component.  The latter provides a better insight into the physical origins of the polarization \cite{seljak1997signature},  since 
at linear order in perturbation theory, primordial scalar perturbations contribute only to the parity-even $E$ mode polarization, whereas primordial tensor perturbations (i.e., gravitational waves) source both $E$ mode and parity-odd $B$ mode polarization at the time of recombination\footnote{Note that on unmasked, full-sky maps there exists a unique mapping between the two bases, but this is not the case in the presence of map boundaries (due to limited observation ranges, or the need to mask foreground-dominated parts of the sky), which introduce an uncertainty in the $(E,B)$ maps one can construct from measured $(Q,U)$ maps.}.

\begin{figure}[t]
    \centering
    \includegraphics[width=0.4\textwidth]{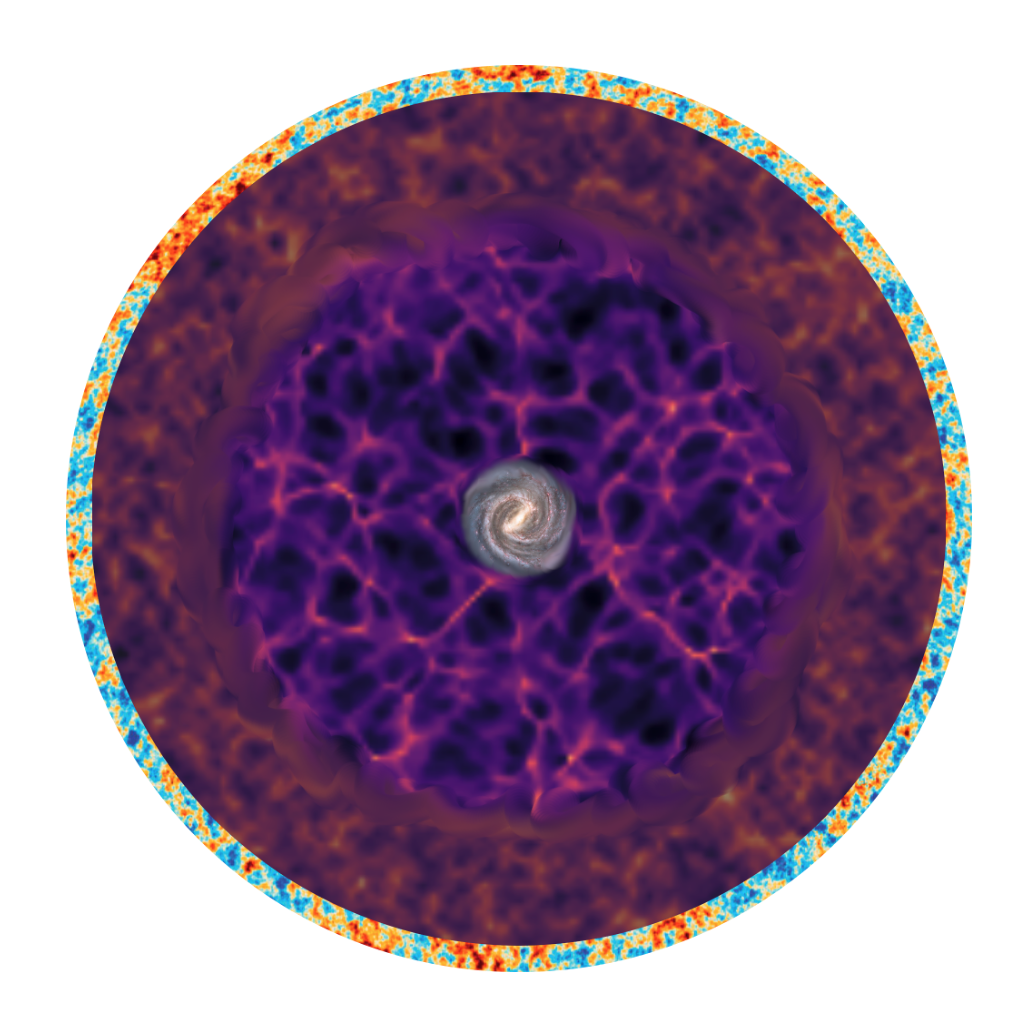}
    \caption{Illustration demonstrating how photons originating from the surface of last scattering, at the outer edges, traverse the large-scale structure of Universe, being lensed in the process, before reaching our galaxy, located at the center. The surface of last scattering refers to the moment in the early Universe when photons, previously trapped in a plasma with electrons and baryons, were first able to travel freely through space, creating the Cosmic Microwave Background we observe today.}
    \label{fig:lensing}
\end{figure}

While purely scalar primordial perturbations imply a vanishing $B$ mode on the last scattering surface, this does not remain true for the CMB we observe today.  This is due to the gravitational lensing effect: on their way from the last scattering surface to Earth, the trajectories of the CMB photons are subject to deflection by the Universe's large scale structure, as sketched in Figure \ref{fig:lensing}.

The deflection can be expressed in terms of the lensing potential $\phi$:
\begin{equation*}
\tilde{X}_{\pm}(\hat{n}) = X_{\pm}(\hat{n} + \nabla \phi(\hat{n})),
\end{equation*}
where $X$ represents the original, unlensed field, and $\tilde{X}$ the lensed field \cite{hu2000weak}, with $X_{\pm} = Q \pm i U$ denoting the polarization fields.  Lensing thus distorts the observed polarization field and results in a transfer of power from the $E$ mode to the $B$ mode.  The presence of this lensing-induced $B$ mode makes the detection of a primordial PGWB-sourced $B$ mode harder.


\subsection{The CMB Data Simulation Process and Dataset}
We assume an underlying cosmology given by the \textit{Planck} best-fit to the base $\Lambda$CDM model~\cite{Planck:2018nkj} plus a contribution from primordial tensor fluctuations with a scale-invariant power spectrum, whose amplitude is parameterized in terms of the tensor-to-scalar ratio $r$. For a given value of $r$,  we calculate the corresponding temperature, polarization and lensing potential angular power spectra using the Boltzmann code \texttt{CAMB}\footnote{\url{https://github.com/cmbant/camb}}~\cite{Lewis:1999bs}. We then use these angular power spectra as input to \texttt{HEALPix}\footnote{\url{https://github.com/healpy/healpy}}~\cite{Gorski:2004by,Zonca2019} to generate random realizations of unlensed full-sky maps of the Stokes $Q$ and $U$ parameters, the $B$ mode polarization and the lensing potential at \texttt{HEALPix} resolution $N_\mathrm{side} = 2048$.  In the next step, we employ the lensing code \texttt{lenspyx}\footnote{\url{https://github.com/carronj/lenspyx}}~\cite{2020ascl.soft10010C} to generate the corresponding lensed full-sky $Q$ and $U$ maps.
Finally, we divide each full-sky map into non-overlapping patches of $12 \times 12$ square degrees and project them onto a $224 \times 224$ pixel square in the Euclidean plane using Cartesian projections about the centre point of each patch. This corresponds to an angular resolution of approximately 3~arcmin, or multipoles $\ell \sim 4000$ in harmonic space – roughly the point at which polarized foregrounds can be expected to start overtaking the CMB signal~\cite{CMB-S4:2022ght}.

Note that CMB maps from realistic experiments would cover a larger area (and would thus be poorly approximated by a planar projection), need to be masked to remove foreground-dominated regions and point sources, and are subject to (potentially anisotropic) noise. Including these effects is technically straightforward, but our goal here being to demonstrate a proof of principle for our method, we opt to consider a simpler, more idealistic scenario for convenience and in order to keep computational requirements manageable.


To generate the dataset for training the neural network, we uniformly sampled 180 values of $r$  within the range of $0$ to $0.3$, motivated by an expected sensitivity to $r$ of roughly 0.1, given our map size and resolution.
For every $r$, we create 20 random realizations of full-sky lensed $Q$ and $U$ maps as well as the corresponding unlensed $B$ maps for training, plus a further 2 realizations for testing/validation purposes.  As described above, each of these full-sky maps is then broken down into 48 independent patches. Our training data thus comprise a total of $48\times 180 \times 20 = 172,800$ $(Q,U,B)$ image sets with corresponding labels of $r$.

\section{Delensing with Variational Encoder-Decoder Algorithm \label{sec:VAE}}

We employ the variational encoder-decoder framework of  \cite{kingma2013auto} to remove the lensing effect, and obtain an unlensed $B$ map with the $Q$ and $U$ maps as input.  We use the variational inference method that maximizes the likelihood of data by approximating the posterior distribution $p(z|Q,U)$ of latent variables $z$, with $q_{\phi}(z|Q,U)$, where $\phi$ denotes the parameters of a neural network. With the approximate $p(z|Q,U)$, we then obtain a posterior predictive distribution for $B$, reconstructing the unlensed $B$ maps from the observed lensed polarization $Q$ and $U$ maps. Mathematically the log-likelihood of $B$ can be written as
\begin{align}\label{eq:ELBO}
\log{p_\theta}(B \mid Q, U) &= \log \int p_\theta(B, z \mid Q, U) d z \nonumber\\
&= \log \int p_\theta(B, z \mid Q, U) \frac{q_\phi(z \mid Q, U)}{q_\phi(z \mid Q, U)} d z \nonumber\\
&= \log \mathbb{E}_{z \sim q_{\phi}{(z \mid Q, U)}}\left[\frac{p_{\theta}(B, z \mid Q, U)}{q_{\phi}(z \mid Q, U)}\right] \nonumber\\
&\geqslant \mathbb{E}_z\left[\log \frac{p_\theta(B, z \mid Q, U)}{q_\phi(z \mid  Q, U)}\right] \nonumber\\
&= \mathbb{E}_z\left[\log p_\theta(B \mid Q, U, z) \right] -D_{\rm KL}\left(q_\phi(z \mid Q, U) \parallel p(z)\right) 
\end{align}
where $q_\phi$, $p_\theta$ represent the probabilistic encoder and decoder respectively, with corresponding network parameters $\phi$ and $\theta$. The function $p(z)$ is the prior distribution of the lower dimensional latent variables $z$ which encodes information from $Q$ and $U$ maps, and $D_{\rm KL}$ denotes the Kullback–Leibler divergence~\cite{Kullback:1951zyt}. The above inequality is also known as evidence lower bound (ELBO) \cite{blei2017}. 

\paragraph{Probabilistic Encoder and Decoder by Swin Transformer}
We adopt a U-Net architecture that incorporates skip connections between the encoder ($q_{\phi}$) and decoder ($p_{\theta}$).  Both $q_{\phi}$ and  $p_{\theta}$ are taken as Gaussian variables. The U-Net architecture benefits from these skip connections as they help preserve the fine grained details of the input polarization maps. The encoder captures the high level contextual information of the input $Q$, $U$ maps, while the decoder reconstructs the unlensed $B$ map from the encoded representation. As the decoder upsamples the feature maps, the spatial resolution increases, but the feature maps might lack the necessary detail to generate an accurate $B$ map. The skip connections provide the decoder with these finer details by directly combining feature maps from the encoder~\cite{caldeira2019deepcmb,yi2020cosmo}.

\begin{figure}
    \centering
    \includegraphics[width=\textwidth]{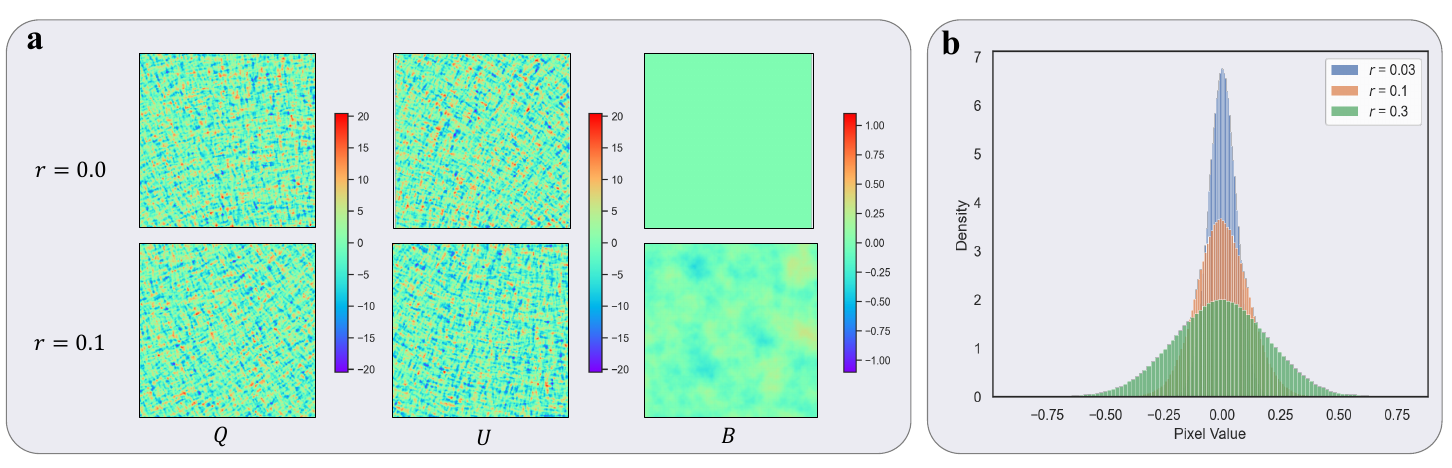}
    \caption{(a) Small patch of $Q$, $U$, $B$ maps for $r = 0.0$ (top row) and $r = 0.1$ (bottom row). The neural network takes $Q$ and $U$ map (left two columns) as input and predicts the $B$ unlensed map (third column). (b) Illustration of the heteroscedasticity 
    of unlensed $B$ maps, where the pixel mean is zero and the variance proportional to $r$.}
    \label{fig:lensing_patch}
\end{figure}

We employ Swin transformer~\cite{liu2021swin} blocks as the core components of both the probabilistic encoder and decoder. The encoder consists of $4$ Swin transformer blocks and is designed to learn the mean and variance of the lower dimensional latent variable $z$ from the high dimensional input data ($Q, U$). The decoder, sharing a similar architecture with the encoder and also comprising 4 Swin transformer blocks, utilizes the latent variables $z$ as input to generate a prediction of the high dimensional unlensed $B$ map.
Additional details of the neural network is given in the Appendix.
Examples of the input/output images in the neural network is depicted in Figure \ref{fig:lensing_patch}(a).

\subsection{Loss Function \label{sec:loss}}
The ELBO  in Equation (\ref{eq:ELBO}) consists of a reconstruction loss and the KL divergence between the prior distribution of $z$ and the learned posterior distribution of $p(z \mid Q,U)$. Maximizing the log-likelihood function $\log p_{\theta}(B \mid Q,U)$ is equivalent to minimizing the negative lower bound with respect to the parameters $\theta$ and $\phi$, then our loss function is:
$$\mathbb{L}=\mathbb{L}_{\mathrm{rec}}(B,\hat{B})+ D_{\rm KL}\left(q_\phi(z \mid Q, U) \parallel p(z)\right) $$
where $\hat{B}$ is the predicted value of $B$.

\paragraph{Reconstruction Loss} The first term of the ELBO is the reconstruction loss of the unlensed $B$ map. We measure the $L_2$ reconstruction loss between the predicted $\hat{B}$ map and the true $B$ map in both pixel space and the power spectra space. For statistically isotropic Gaussian fields, such as the CMB anisotropies, the power spectra encapsulate the full statistical information and are thus commonly used in the construction of likelihood functions in CMB data analysis.
Specifically, the amplitude of the unlensed $B$ mode polarization power spectrum is directly proportional to $r$, making it a particularly useful probe of the PGWB.
Combining the $L_2$ loss in pixel space and the $L_2$ loss in the power spectrum domain ensures that our model effectively captures the essential features of the unlensed $B$ map.

A na{\"i}ve definition of the reconstruction loss $\mathbb{L}_\text{rec}$ is as follows:

$$\mathbb{L}_\text{rec} = \Bigl\|B-\hat{B}\Bigr\|_2^2 + \alpha \, \Bigl\| P(k) -  \hat{P}(k)\Bigr\|_2^2,$$
where $\hat{P}(k)=  \frac{1}{N(k)} \sum_{k_x, k_y \in \text{bin}(k)} \left|F_{k_x k_y}^B\right|^2$, and $F_{k_x k_y}^B$ is the Fourier transform of the unlensed $B$ map. Here, $N(k)$ denotes the number of $(k_x, k_y)$ pairs within the radial bin corresponding to the wave vector magnitude $k$. $k_x,k_y$ are the wave components, and $\alpha>0 $ is a constant that determines the relative contributions of the pixel space and power spectrum terms.  

However, in practice the above definition is problematic since the
unlensed $B$ maps are heteroscedastic: while the mean of the pixel values is zero, their variance is proportional to the parameter $r$, as illustrated in Figure \ref{fig:lensing_patch}(b).
As a consequence, the same relative prediction error would result in a much higher loss for large values of $r$ than it would for small values, causing the model to prioritize minimizing errors for maps with larger $r$ values.  This would lead to biased inference and 
poorer performance for maps with smaller $r$ values.
Similarly, since we are considering noiseless maps, the power spectrum $P(k)$ displays heteroskedastic behaviour as well, with its variance proportional to its amplitude, which in turn is proportional to $r$.

To tackle these issues, we scale the pixel domain loss with a factor of $1/r$ and, instead of the difference in power spectra, we consider the difference in the logarithm of the power spectra, whose variance is independent of $r$.
This approach ensures that all values of $r$ are penalized equally, allowing us to fit the neural network 
across maps with varying $r$ values.   Our final reconstruction loss is therefore given as:

\begin{equation}\label{eq: loss function}
    \mathbb{L}_\text{rec} = \Bigl\|\frac{1}{\sqrt{r}}B-\frac{1}{\sqrt{r}}\hat{B}\Bigr\|^2_2 + \alpha \, \Bigl\| \log P_{r}(k) -  \log \hat{P}_{r}(k)\Bigr\|_2^2.
\end{equation}
For the relative weighting factor $\alpha$, we found that values around $0.3$ work well.

As the available information for predicting the edges of the unlensed $B$ map from the cropped $Q$ and $U$ maps is inherently limited, the edges of the predicted $B$ maps invariably default to the mean value of the $B$ maps, typically zero. This phenomenon is similar to those results predicted by the likelihood and neural network-based methods~\cite{caldeira2019deepcmb,millea2019bayesian}. Consequently, the loss function for the edge regions manifests as an irreducible loss. As we scale the reconstruction loss to address the heteroscedasticity, scaling the irreducible loss in the edge region will also lead to bias optimization for different $r$. To address this concern, we restrict our reconstruction loss calculations to the central $180 \times 180$ region of the $B$ maps. This approach ensures that our predicted $B$ maps are unbiased across all values of $r$.

\paragraph{KL Divergence} 
The second component of the loss function is the regularization term, which quantifies the divergence between the posterior distribution $p(z|Q,U)$ and the prior distribution $p(z)$. We define the prior distribution $p(z)$ as a standard Gaussian distribution $\mathcal{N}(0, I_D)$ and model the posterior distribution $p(z|Q,U)$ as a Gaussian distribution $\mathcal{N}(\mu, \sigma^2I_D)$, where $\mu$ and $\sigma^2$ are parameters estimated by the encoder network $q_\phi$, $D$ is the dimension of the latent space, and $I_D$ is a $D\times D$ dimensional diagonal matrix. Given that both distributions are Gaussian, the KL divergence can be computed analytically as follows:

$$D_{\mathrm{KL}}(q(z|Q,U) || p(z)) = \frac{1}{2} \sum_{i=1}^D \left( \sigma_i^2 + \mu_i^2 - 1 - \log(\sigma_i^2) \right).$$

In this equation,  $\mu_i$ and $\sigma_i^2$ are the mean and variance of the $i$-th latent variable estimated by the encoder.  Here we set $D = 100$. 

\subsection{Optimization of the Neural Network}

During training, we adopt a straightforward approach by randomly selecting a full-sky map and using all its 48 cropped segments as a single batch. With the loss function and sampling strategy above, we utilized the Adam optimizer to train the whole neural network with initial learning rate of $10^{-5}$. If there is no improvement in the model's performance on the validation dataset for 10 consecutive epochs, we decrease the learning rate by a factor of 0.1. We stop training the model after 230 epochs, with no further improvement in validation loss. The training process is carried out on an NVIDIA V100 GPU.

\section{Approximate Bayesian Inference for Tensor-to-Scalar Ratio $r$ \label{sec:abcinference}}

\begin{figure}[t]
    \centering
    \includegraphics[width=\textwidth]{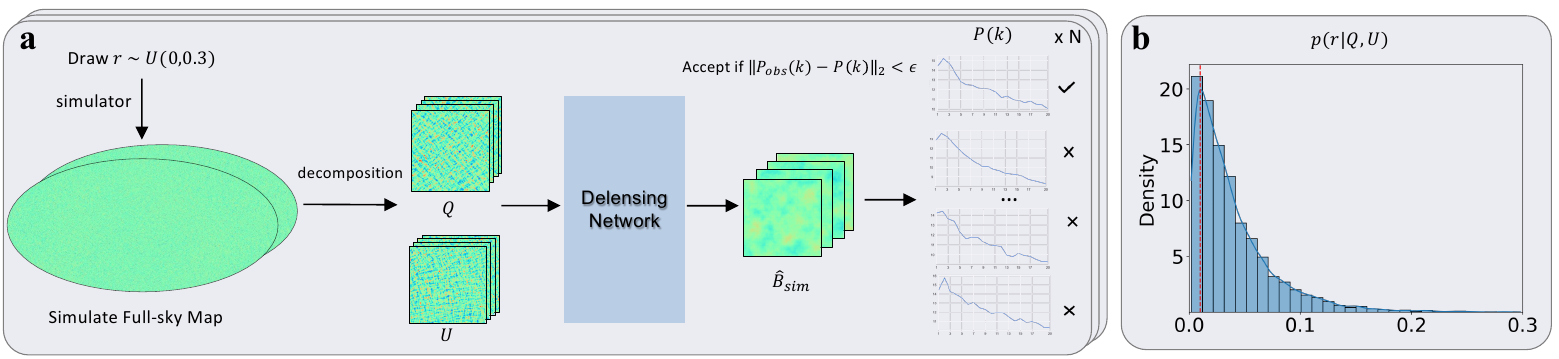}
    \caption{Overview of simulation-based inference for $r$. (a) outlines the initial steps of our inference process, starting with sampling $r$ to generate full-sky lensed $Q$ and $U$ maps. These maps are segmented into smaller square patches, from which we predict the corresponding $B$ unlensed map and using its power spectrum $P(k)$ as the summary statistics of the data, as part of the ABC algorithm to produce the posterior for $r$.  (b) displays the aggregation of accepted $r$ samples, which forms the final posterior distribution $p(r|Q_\mathrm{obs},U_\mathrm{obs})$.}
    \label{fig:all process}
\end{figure}

With a neural network-assisted approach for delensing and dimension reduction via power spectrum, we can now apply the ABC algorithm  to infer the tensor-to-scalar ratio $r$. We shall dub our approach the ``AB$\mathbb{C}$MB algorithm'' – described in detail in Algorithm~2, and a pictorial depiction of the algorithm flow can be found in Figure~\ref{fig:all process}.
Unlike traditional MCMC methods, which may be computationally intensive in high-dimensional data settings, ABC circumvents the need to evaluate the likelihood function $L(Q,U | r)$, but instead simulates the data $Q$ and $U$ given $r$. The discrepancy between simulated and observed data is computed, with the discrepancy approaching zero indicating that the given $r$ should be retained as a sample from the posterior. 
Here, using the discrepancy between the simulated unlensed $B_n$ map with the unlensed $B_\mathrm{obs}$ map from the observed data $Q_\mathrm{obs}, U_\mathrm{obs}$ would be more informative for $r$ than working directly with $Q_\mathrm{obs}, U_\mathrm{obs}$.

\begin{minipage}{0.95\textwidth}
    \begin{algorithm}[H]
        \caption{AB$\mathbb{C}$MB: Deep learning assisted ABC algorithm}  
        \begin{algorithmic}[1]
            \For{$n = 1 : N$}
                \State Sample $r_n \sim U(0, 0.3)$.
                \State Simulate maps $Q_n, U_n \sim p(Q, U | r_n)$ using the simulator $p(Q, U|r_n)$.
                \State Obtain delensed $B_n$ map:
sample $z_j\sim q_{\phi}(z|Q_n,U_n)$ and $B_j \sim p_{\theta}(B|z_j)$, set $B_n = \sum_{j=1}^J B_j/J.$
                \State Compute summary statistics $S(Q_n, U_n) =P_n(k) $ as power spectrum  of unlensed $B_n$ map, where $P_n(k)=  \frac{1}{N(k)} \sum_{k_x, k_y \in \text{bin}(k)} \left|F_{k_x k_y}^B\right|^2$, and $F_{k_x k_y}^B$ is the Fourier transform of the $B_n$.
                \State Accept $r_n$ if $\|P_n(k) - P_\mathrm{obs}(k)  \|_2  < \epsilon$.
            \EndFor
       \end{algorithmic}  
        \label{vABC algorithm}
    \end{algorithm}
\end{minipage}

\paragraph{Summary Statistics} 
Since the CMB problem involves over 100,000 dimensions in our example, reducing the discrepancy to below an arbitrarily small tolerance $\epsilon$ would be computationally prohibitive.  However, it has been shown that dimension reduction can be an effective way to improve the performance of the algorithm \cite{sissonyb2018}. Since CMB is an isotropic Gaussian random field, the power spectra encapsulate the full statistical information without any loss. Given that $r$ is defined as the ratio of the amplitudes of the tensor to scalar power spectra, $r \equiv A_\mathrm{t}/A_\mathrm{s}$, it can summarize the $B$ maps with no loss of information concerning $r$. We therefore compute the discrepancies between the simulated and observed power spectra instead of directly between the $B$ maps.

\paragraph{Approximate posterior} The final posterior distribution is obtained by collecting all $r_n$ values that produce simulated data within a tolerance $\epsilon$ of the observation. For $\epsilon > 0$, an approximation error is introduced. The larger values of $\epsilon$ are computationally cheaper but tend to produce bias and overestimate uncertainty. In the other direction, reducing $\epsilon$ to zero is not attainable in many real world problems, therefore in practice $\epsilon$ is chosen either at a computational limit or when the improvement in the posterior is no longer noticeable. An alternative to directly specifying the $\epsilon$ value is to simulate a large number of realisations, and keep a small percentage (e.g., 5\%) of samples with the smallest discrepancy, the later is the approach adopted in this paper.

Formally, our approximated posterior distribution using Algorithm 2 takes the form
$$
 p(r|Q_\mathrm{obs},U_\mathrm{obs}) \approx p_\epsilon(r|Q_\mathrm{obs},U_\mathrm{obs}) \propto  p(r) \mathbb{I}(\|P_n(k) - P_\mathrm{obs}(k)  \|_2  < \epsilon),
$$
where $\mathbb{I(\cdot)}$ takes the value 1 if the condition is satisfied, and 0 otherwise.

\section{Results \label{sec:results}}
In this section, we will show both delensing and posterior estimation results. Firstly, we evaluate our predicted unlensed $B$ polarization map in pixel and power spectrum spaces. To test the unbiasedness of our predictions, we compare the mean predicted power spectrum across 48 realizations (comprising all patches in the entire sky map) with the mean power spectrum of the true unlensed $B$ mode map. Furthermore, we leverage the power spectrum as a summary statistic to estimate the posterior distribution of the tensor-to-scalar ratio $r$ by ABC. To validate the robustness of our approach, we also conduct further experiments to vaildate that our posterior has the correct coverage  properties in the frequentist sense.

\subsection{Delensing}
To evaluate the 
quality of our model prediction, we compare each sample and the mean of the estimated unlensed $B$ map with ground truth in both pixel and power spectrum space. 

\begin{figure}[t]
    \centering
    \begin{subfigure}{0.24\textwidth}
        \centering
        \includegraphics[width=\textwidth]{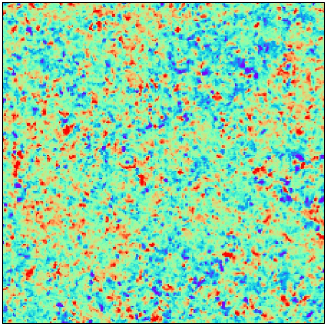}
        \caption{Lensed $B$ Map }
    \end{subfigure}%
    \vspace{1mm}
    \begin{subfigure}{0.24\textwidth}
        \centering
        \includegraphics[width=\textwidth]{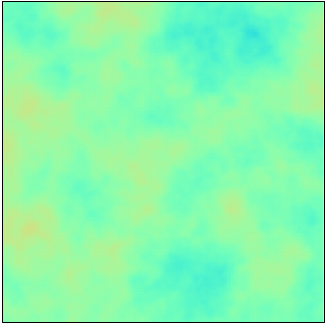}
        \caption{Unlensed $B$ Map}
    \end{subfigure}
    \begin{subfigure}{0.24\textwidth}
        \centering
        \includegraphics[width=\textwidth]{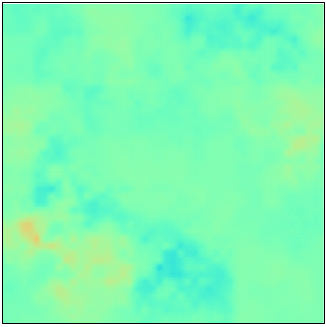}
        \caption{AB$\mathbb{C}$MB}
    \end{subfigure}    
    \begin{subfigure}{0.24\textwidth}
        \centering
        \includegraphics[width=\textwidth]{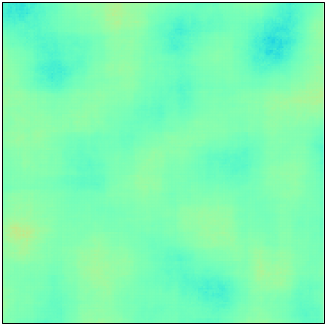}
        \caption{DeepCMB}
    \end{subfigure}     
    \caption{(a) Simulated lensed $B$ map , (b) The true unlensed $B$ mode map, serving as the baseline for comparison. (c) Our model predicted unlensed $B$ map. We used the mean of posterior $p(z|Q,U)$ and compute $p(B|z)$. (d) DeepCMB method's prediction. }
    \label{fig:comparison between samples and truth}
\end{figure}

\paragraph{Pixel Space}
In the pixel space domain, we estimate the unlensed $B$ map by calculating the posterior mean of $p(z|Q,U)$, which is then input into a decoder to generate the unlensed $B$-mode maps. We compare this estimation with the ground truth and the results from the DeepCMB method, as shown in Figure~\ref{fig:comparison between samples and truth}. Our analysis reveals that our model captures a higher level of detail in the unlensed $B$-mode maps compared to deterministic methods like DeepCMB~\cite{caldeira2019deepcmb}.

\paragraph{Power Spectrum}
Beyond the comparison in pixel space, we further assess our model in the context of the power spectrum. The power spectrum serves as a pivotal tool for the analysis of CMB maps. It is particularly significant when it comes to the tensor-to-scalar ratio $r$, as the power spectrum is linearly proportional to $r$. 

\begin{figure}[!htp]
    \centering
    \begin{subfigure}{0.5\textwidth}
        \centering
        \includegraphics[width=\textwidth]{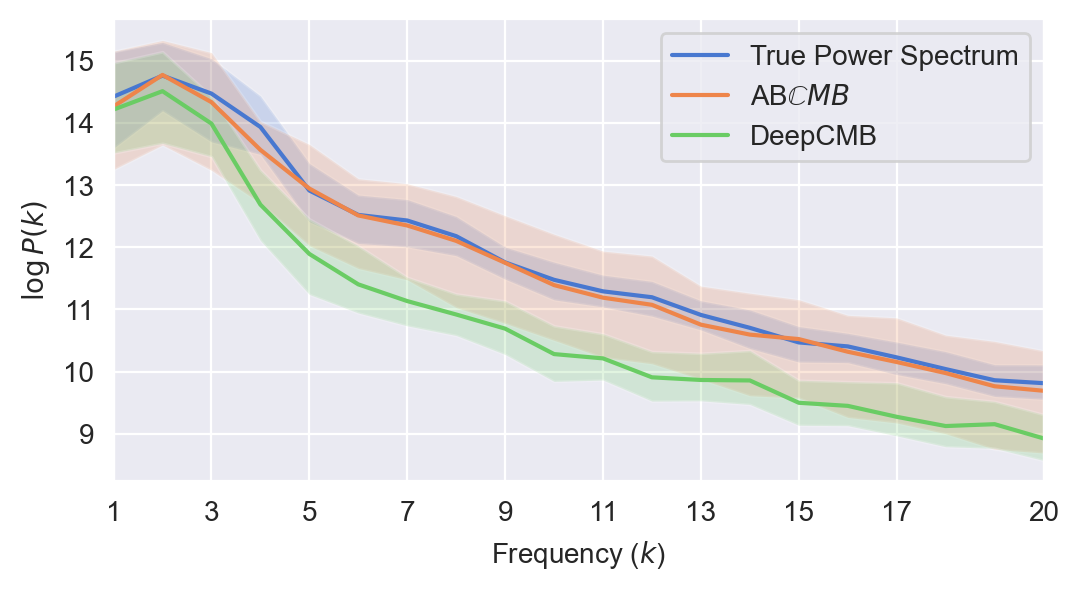}
    \end{subfigure}%
    \hfill
    \begin{subfigure}{0.5\textwidth}
        \centering
        \includegraphics[width=\textwidth]{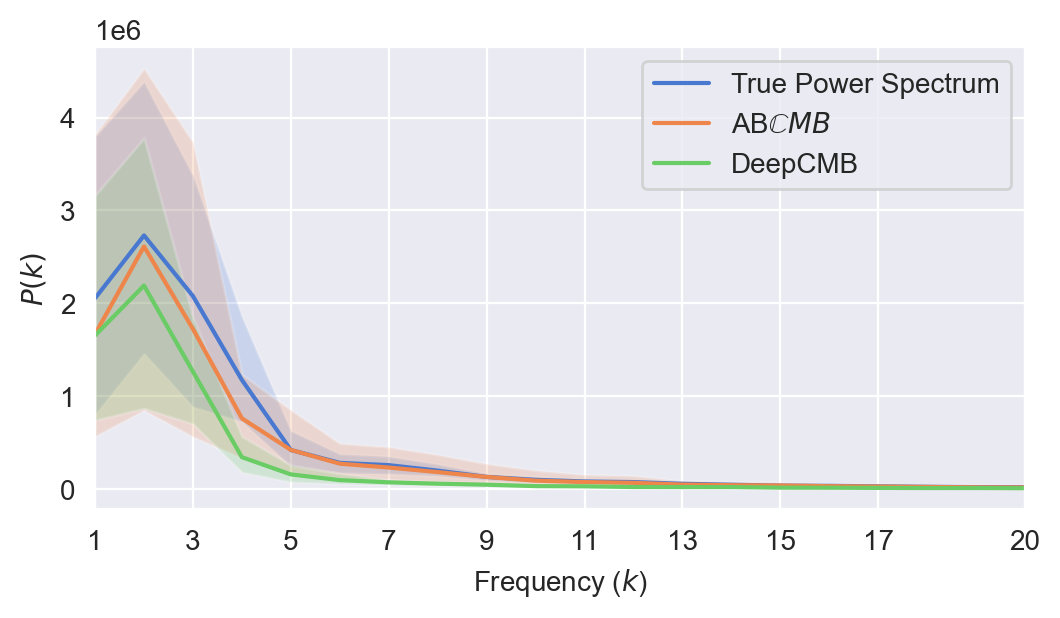}
    \end{subfigure}
    \caption{Predicted power spectrum for $r = 0.1$ using AB$\mathbb{C}$MB (orange), DeepCMB (green) and the ground truth (blue). Shaded areas correpond to their respective 0.05 and 0.95 percentile estimates from the 48 patches.
    Displayed in logarithmic (left) and linear (right) scales.
    \label{fig:comparison ps between samples and truth}}
\end{figure}

To 
illustrate the unbiasedness of our model at the power spectrum level, we calculate the mean power spectrum of 48 predicted $B$ maps and compare this with the ground truth. As shown in Figure \ref{fig:comparison ps between samples and truth}, the mean of the predicted power spectrum aligns closely with the true power spectrum, even at high frequencies. 

\subsection{Posterior Distribution Results}
\begin{figure}
    \centering
    \includegraphics[width=0.99\textwidth]{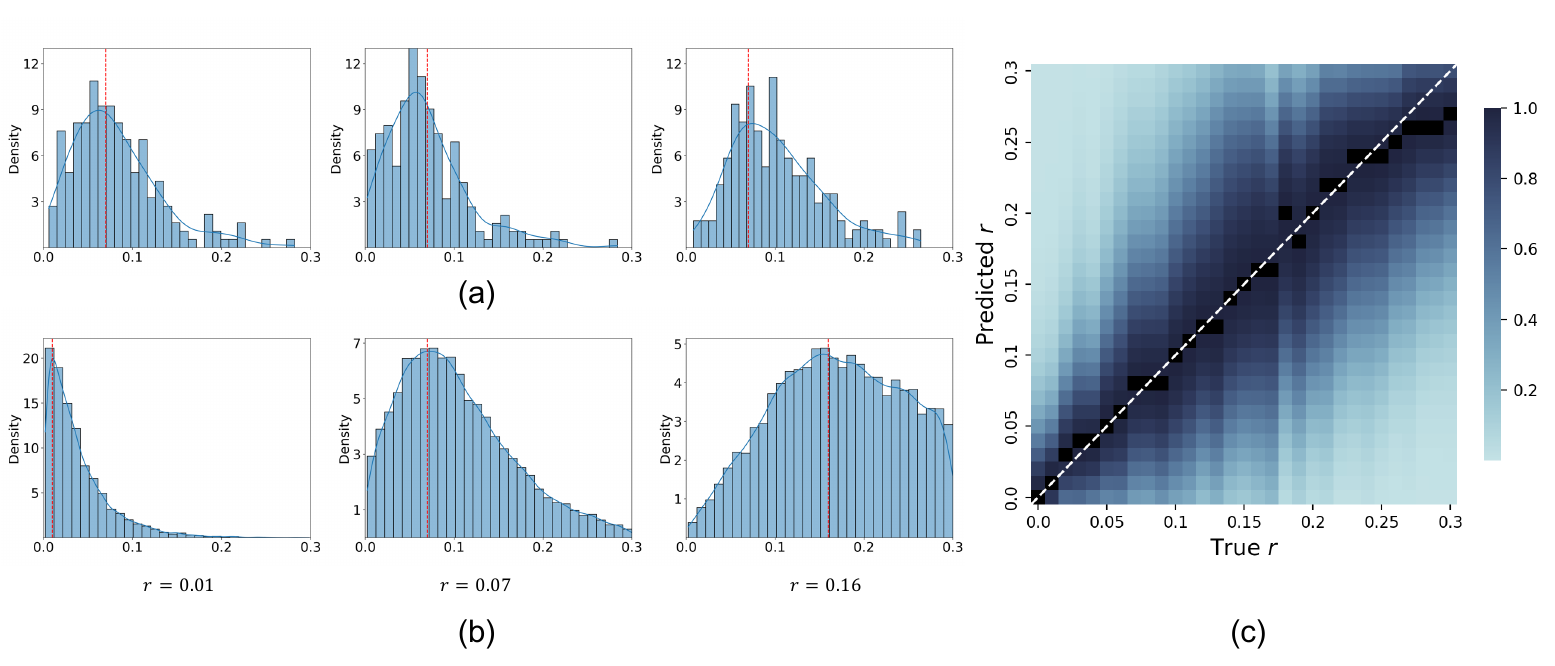}
    \caption{(a) Posterior distribution inferred from three patches of the CMB map which corresponding $r$ is $0.07$. (b) Posterior distribution for $r$ after aggregating 48 patches. (c) Posterior distribution for $r$ ranging from $0$ to $0.3$. Vertical line indicate the true value of $r$. The black block indicates the mode of posterior on each column.}
    \label{fig: posterior result}
\end{figure}

By accurately and unbiasedly estimating the power spectrum and using it as a summary statistic, we can then use ABC to estimate the tensor-to-scalar ratio given high dimensional $Q$ and $U$ maps. We sample $r$ from a uniform prior $p(r) \sim U(0, 0.3)$ and simulate $ Q, U $ maps using the simulator for each $r$. Subsequently, we predict the unlensed $ B $ maps by neural networks for each simulated lensed map and compute their corresponding power spectra as summary statistics. Out of 40,000 sampled maps, we accept the corresponding values of $r$ for which the computed power spectra exhibit the smallest $5\%$ of errors relative to the predicted $B$ map's power spectrum for observed data. As shown in Figure \ref{fig: posterior result}(a), 2000 samples of $r$ are obtained for each observed small patch $Q, U$ maps, to form the posterior $p(r|Q, U)$.  


To assess whether the posterior is well calibrated, that is, that the posterior credibility interval has the correct coverage in the frequentist sense. By this we mean, that for nominal 95\% credibility interval, out of multiple posteriors obtained from multiple realizations of the data, approximately 95\% of these intervals would contain the true $r$.  To check this, we computed the empirical coverage probabilities.  Specifically, for each posterior distribution, we calculate the 95\% high-density posterior region and evaluate whether the ground truth value falls within this interval. This process is repeated multiple times to ascertain the frequency with which this high-density region captures the true value. Theoretically, the proportion of the true value of $r$ residing within the high-density posterior region should equal the stated credibility level. To validate this, we utilize $r = 0.07 $ as a test case and sampled 96 distinct maps, resulting in 96 different posterior credible intervals. Remarkably, 92 out of these 96 credibility intervals contain the true value of $r$, an empirical coverage of 95.5\%, demonstrating a high level of accuracy in the ABC posterior.






For a single realization of the data, the inferred posterior is subject to considerable sample variance.  This effect can be suppressed by aggregating posteriors from multiple patches – we use this to demonstrate that when averaged over many realizations of the data, our method correctly recovers the input value of $r$.
Aggregating data from multiple patches can significantly reduce the uncertainty estimate. To achieve this, we simply aggregated the 2000 accepted samples from all 48 patches and to estimate the final posterior distribution. As illustrated in Figure \ref{fig: posterior result}(b), the posterior distribution based on 48 maps is demonstrably more reliable than that based on single samples. For both small and large values of $ r $, the posterior distribution centers around the true tensor-to-scalar ratio. The larger the number of small patches observed, the more precise the estimated posterior distribution becomes. To comprehensively evaluate this phenomenon, we examined the posterior distributions of 31 different $r$ values, ranging from 0 to 0.3, as shown in Figure \ref{fig: posterior result}(c). For each $r$ value, we computed its respective posterior distribution, normalizing it by its maximum amplitude. The resulting heatmap displays a diagonal pattern, indicating that our methodology yields posteriors of $r$ whose modal values center on the truth. We note that for larger values of $r$, we observe an increase in the uncertainty. This is to be expected: as discussed in Section \ref{sec:loss} the variance of $r$ is proportional to $r$. For values of $r$ closer to the upper cutoff boundary of 0.3, we can see that the results are a little underestimated. Thus extending the range on the prior might be needed if the true $r$ is higher.





\section{Conclusion and Discussion \label{sec:conclusions}}
In this paper, we have developed a likelihood-free inference method to deduce the posterior distribution of the tensor-to-scalar ratio from lensed CMB polarization maps, integrating 
knowledge of the underlying physics
into our approach through the use of deep generative models for delensing as an intermediate step. Our methodology yields unbiased delensing results, utilizing the power spectrum of delensed $B$ maps as summary statistics for inferring $r$. The resulting posterior distribution provides unbiased estimates of $r$ with well-calibrated uncertainty quantification.

A significant advantage of the proposed approach is its independence from traditional likelihood formulations, eliminating the need for construction of likelihood functions based on lensing potential and the tensor-to-scalar ratio. Additionally, the proposed method is computationally efficient, with the entire process being fully parallelizable. 
We have demonstrated that our method works reliably and efficiently on the somewhat idealized example of small, noiseless, square, planar patches of the CMB sky.  A generalization to more realistic CMB observations with larger observed areas, spherical geometry, additional, potentially anisotropic, noise contributions, and non-trivial survey geometry due to masking is straightforward: for any given experiment, these effects are known and can simply be included at the data simulation stage.

Future work should investigate the performance of the method with real CMB data, under more complex settings.


\section*{Acknowledgements}
    Some of the results in this paper have been derived using the \texttt{healpy} and \texttt{HEALPix} packages.  This project was undertaken with the assistance of computational resources and services from the National Computational Infrastructure (NCI), which is supported by the Australian Government.


\bibliographystyle{unsrt}
\bibliography{sample}

\newpage
\appendix
\section*{Appendix: Neural Network}
In this study, we employed the Swin Transformer\cite{liu2021swin} as the primary neural network architecture for both the probabilistic encoder and decoder within the VAE framework. The Swin Transformer is a hierarchical neural network designed to learn multiscale information, making it highly suitable for our problem of capturing spatial correlations and structures in the CMB maps. The Swin Transformer utilizes local windows to capture contextual information at multiple scales and leverages hierarchical processing to efficiently model spatial correlations and structures in the input data.
\paragraph{Attention Mechanism}

The foundation of the Vision Transformer\cite{dosovitskiy2021an} is the multi-head self attention (MSA) mechanism, which computes the self-attention operation for a given input feature map. The MSA operation can be defined as:

\begin{equation}
\mathrm{MSA}(X) = \mathrm{SoftMax}\left(\frac{QK^T}{\sqrt{d_k}}\right)V,
\end{equation}
where $Q$, $K$, and $V$ are the query, key, and value matrices obtained by first applying a linear transformation to the input feature map $X$, $d_k$ is the dimension of the keys. For example, the query matrix $Q$ is computed as follows:

\begin{equation}
Q = XW + b,
\end{equation}
where $W$ and $b$ are the learnable weight matrix and bias term  The key matrix $K$ and value matrix $V$ can be computed similarly, by the different trainable parameters $W$ and $b$, respectively. 

\paragraph{Swin Transformer Block}
The Swin Transformer block incorporates the window-based multi-head self attention (W-MSA) module and the Shifted window-based multi-head self attention (SW-MSA) module, which enables the model to efficiently capture local information and global context.

 
 The W-MSA module processes the input feature map by dividing it into non-overlapping windows of a fixed size. Within each window, the standard MSA operation is performed independently. By restricting the attention computation to local windows, the W-MSA module significantly reduces the computational complexity of the attention mechanism, making it more efficient and scalable for larger input sizes. The SW-MSA module builds upon the W-MSA module by introducing a shift operation before applying the MSA mechanism. The input feature map is first divided into non-overlapping windows, and each window is then shifted by half of its size in both the vertical and horizontal dimensions. This shifting operation ensures that the model captures both local and global context, as it allows the MSA mechanism to attend to a broader range of spatial positions.
 
By utilizing the shifted window partitioning approach, the consecutive Swin Transformer blocks are computed as follows:
\begin{align}
& \hat{\mathbf{x}}^l=\operatorname{W-MSA}\left(\mathrm{LN}\left(\mathbf{x}^{l-1}\right)\right)+\mathbf{x}^{l-1} \nonumber\\
& \mathbf{x}^l=\operatorname{MLP}\left(\mathrm{LN}\left(\hat{\mathbf{x}}^l\right)\right)+\hat{\mathbf{x}}^l \nonumber\\
& \hat{\mathbf{x}}^{l+1}=\operatorname{SW-MSA}\left(\mathrm{LN}\left(\mathbf{x}^l\right)\right)+\mathbf{x}^l \nonumber\\
& \mathbf{x}^{l+1}=\operatorname{MLP}\left(\mathrm{LN}\left(\hat{\mathbf{x}}^{l+1}\right)\right)+\hat{\mathbf{x}}^{l+1}\nonumber
\end{align}
where $\hat{\mathbf{x}}^l$ and $\mathbf{x}^l$ denote the output features of the (S)WMSA module and the MLP module for block $l$, respectively. $\mathrm{LN}$ denotes layer normalisation\cite{ba2016layer} and $\operatorname{MLP}$ is multi-layer perception.
The detail neural network is :

These encoder and decoder architectures:
$$\begin{aligned} \{Q,U\} \in \mathcal{R}^{224 \times 224 \times 2} 
& \stackrel{\operatorname{Patch Partition}}{\longrightarrow} \{Q,U\}_{\text{patch}} \in \mathcal{R}^{56 \times 56 \times 32}   \\ 
 & \stackrel{\operatorname{Swin Block}}{\longrightarrow} \Psi_1 \in \mathcal{R}^{56 \times 56 \times 96}   \\ &\stackrel{\operatorname{Swin Block}}{\longrightarrow} \Psi_2 \in \mathcal{R}^{28 \times 28 \times 192}    \\ & \stackrel{\operatorname{Swin Block}}{\longrightarrow} \Psi_3 \in \mathcal{R}^{14 \times 14 \times 384}    \\& \stackrel{\operatorname{Swin Block}}{\longrightarrow} \Psi_4 \in \mathcal{R}^{7 \times 7 \times 768}    \\& 
 \stackrel{\operatorname{Flatten}}{\longrightarrow} \Psi_5 \in \mathcal{R}^{37632}   \rightarrow \mathrm{FC}_{100} 
\\& \rightarrow \mu \in  \mathcal{R}^{100} ,\sigma^2 \in \mathcal{R}^{100} \\& \rightarrow z = \mu + \sigma \odot \epsilon \in \mathcal{R}^{100}\\
\\z \in \mathcal{R}^{100} & \longrightarrow \mathrm{FC}_{37632} \rightarrow  \Phi_4 \in \mathcal{R}^{7 \times 7 \times 768}\\ & \stackrel{\operatorname{Swin Block}}{\longrightarrow} (\Phi_3 + \Psi_3) \in \mathcal{R}^{14 \times 14 \times 768}  \\& \stackrel{\operatorname{Swin Block}}{\longrightarrow} (\Phi_2 + \Psi_2) \in \mathcal{R}^{28 \times 28 \times 384} \\ & \stackrel{\operatorname{Swin Block}}{\longrightarrow} (\Phi_1 + \Psi_1) \in \mathcal{R}^{56 \times 56 \times 192} 
\\ &  \stackrel{\operatorname{Swin Block}}{\longrightarrow}  \operatorname{Patch Merge} \rightarrow B \in \mathcal{R}^{224 \times 224 \times 1},\end{aligned}
$$
\end{document}